\documentclass[titlepage,prb,aps,groupedaddress,preprint,12pt,nobibnotes]{revtex4}

\usepackage[dvips]{graphicx}%

\begin{document}

\title{Interpretation of the ``S-shaped'' temperature dependence of
  luminescent peaks from semiconductors}


\author{Q.~Li}
\affiliation{Department of Physics and HKU-CAS Joint Laboratory on New
  Materials, The University of Hong Kong, Pokfulam Road, Hong Kong,
  China.}

\author{S.~J.~Xu}%
\thanks{Author to whom correspondence should be
    addressed; Electronic mail: sjxu@hkucc.hku.hk.}
\affiliation{Department of Physics and HKU-CAS Joint Laboratory on New
  Materials, The University of Hong Kong, Pokfulam Road, Hong Kong,
  China.}

\author{M.~H.~Xie}
\affiliation{Department of Physics and HKU-CAS Joint Laboratory on New
Materials, The University of Hong Kong, Pokfulam Road, Hong Kong, China.}

\author{S.~Y.~Tong}%
\thanks{Present address: Department of Physics and
    Materials Science, City University of Hong Kong, Kowloon Tang,
    Hong Kong, China.}%
\affiliation{Department of Physics and HKU-CAS Joint Laboratory on New
  Materials, The University of Hong Kong, Pokfulam Road, Hong Kong,
  China.}



\begin{abstract}

The ``S-shape'' (decrease-increase-decrease) temperature dependence of
luminescence peak shift from semiconductors is considered. A
luminescence model for localized state ensemble was employed to
interpret this anomalous temperature dependence of emission
peak. Excellent agreement between the theoretical calculation and the
experiments was achieved over the whole studied temperature
region. The physical origin of the ``S-shaped'' shift is revealed.

\end{abstract}


\maketitle


Anomalous temperature dependence of luminescence peak position has
been frequently observed in
semiconductors. \cite{ChoYH1998-APL73:1370, EliseevPG1997-APL71:569,
  GrenouilletL2000-APL76:2241, HongYG2003-APL83:5446,
  YuHB2004-CPL21:1323, BellA2004-JAP95:4670, CaoXA2003-APL82:3614,
  OlsthoornSM1993-JAP73:7798} A typical such phenomenon is so-called
``S-shaped'' temperature dependence. That is, as the temperature
continuously increases, the luminescence peak redshifts first, then
blueshifts, finally redshifts again. This unusual temperature
dependence of luminescence peak significantly deviates from that
predicted by either the Varshni \cite{VarshniYP1967-Physica34:149} or
Bose-Einstein \cite{VinaL1984-PRB30:1979} formula. For an ideally
perfect semiconductor, these formula predict a monotonic decrease of
the semiconductor band gap with increasing temperature. The
``S-shaped'' temperature-induced shift of luminescence peak has been
known to be closely related to the carrier localization in
semiconductors for a long time. \cite{ChoYH1998-APL73:1370} Several
attempts have been made to interpret the phenomenon. For example, a
band-tail emission model was proposed by Eliseev \textit{et al.}
\cite{EliseevPG1997-APL71:569} to explain the frequently observed
blueshift of the emission peak in InGaN QWs at higher
temperatures. However, to our best knowledge, no a model can reproduce
the ``S-shaped'' shift.

In this letter, starting from a newly developed luminescence model for
localized state ensemble, \cite{LiQ2001-APL79:1810,
  LiQ-cond-mat-0411128} we quantitatively interpret the ``S-shaped''
temperature dependence of luminescence peaks from several
representative materials. It is found that the redistribution of
carriers within the localized states due to the transfer between
different localized states and the thermal escape of carriers from the
higher energy states leads to the occurrence of the ``S-shaped''
temperature dependence of the luminescence paek. Excellent agreement
between the theoretical calculation and the experimental data is
achieved over the whole studied temperature range.


Figure~\ref{Fig:S-shape} shows the luminescence peak positions against
temperature for four different materials of InGaAsN,
\cite{GrenouilletL2000-APL76:2241} GaInNP,
\cite{HongYG2003-APL83:5446} InGaN, \cite{YuHB2004-CPL21:1323} and
AlGaN \cite{BellA2004-JAP95:4670}. The solid squares represent the
experimental data while the solid lines are the calculated results
using the model described in detail in
ref. \onlinecite{LiQ-cond-mat-0411128}. The emission colors of these
materials vary from ultraviolet (AlGaN, 3.9 eV) to near-infrared
(InGaAsN, 0.965 eV). In these materials, the emission spectra are from
the radiative recombination of the localized excitons due to either
large-scale inhomogeneity in composition or the N-induced localized
states. \cite{GrenouilletL2000-APL76:2241, HongYG2003-APL83:5446,
  YuHB2004-CPL21:1323, BellA2004-JAP95:4670, JohnsonNM2000-PT53:31}
Considering the radiative recombination, thermal escape, and
re-capture of the excitons in a localized state ensemble, we derive a
distribution function of localized carriers from a rate equation under
quasi-steady state. \cite{LiQ-cond-mat-0411128} Assuming that the
localized state ensemble has a Gaussian-type energetic distribution of
density of states, the luminescence spectrum of the localized excitons
is found. The temperature dependence of the luminescence peak
positions can be given by \cite{LiQ-cond-mat-0411128}
\begin{eqnarray}
   E(T)=E_0-\frac{\alpha T^2}{T+\Theta}-x(T)\cdot k_BT.
   \label{eq:peak-position}%
\end{eqnarray}
where the second term on the right describes the bandgap shrinking
according to Varshni's empirical formula. $\alpha$ is the Varshni's
parameter and $\Theta$ the Debye temperature. The third term
represents the effect of thermal redistribution of localized
carriers. $k_B$ is the Boltzmann constant. The dimensionless
coefficient $x(T)$ can be obtained by numerically solving the
following equation \cite{LiQ-cond-mat-0411128, LiQ2001-APL79:1810}
\begin{eqnarray}
   xe^x=\left[\left(\frac{\sigma}{k_BT}\right)^2-x\right]
   \left(\frac{\tau_r}{\tau_{tr}}\right)e^{(E_0-E_a)/k_BT},
   \label{eq:solve-x}%
\end{eqnarray}%
where $E_0$ and $\sigma$ are the central energy and broadening
parameter for the distribution of the localized states,
respectively. $1/\tau_{tr}$ is the escape rate and $1/\tau_r$ the
radiative recombination rate of the localized carriers. Like the Fermi
level in the Fermi-Dirac distribution function, $E_a$ gives a special
energy level below which the localized states are occupied by the
excitons at 0 K. It is shown that the magnitude and sign of $E_a-E_0$
strongly affect the temperature dependence of the luminescence
paek. When $E_a-E_0$ is taken a negative value, the ``S-shaped''
temperature dependence (the solid lines) of the luminescence peak can
be well produced, as shown in Fig.~\ref{Fig:S-shape}. The parameters
adopted in the calculations are summarized in Table
\ref{Tab:parameters}.

As mentioned earlier, $E_a-E_0$ plays an important role in determining
the temperature behavior of the luminescence peak of the localized
excitons, particularly in the low temperature region. As the central
energetic position of the localized state distribution, $E_0$ is
determined for a given material. However, the energetic position of
$E_a$ is found to be dependent on both the concentration of carriers
and the magnitude of the built-in electric field in the material. The
magnitude and even the sign of $E_a-E_0$ can thus be changed through
internally or externally adjusting the carrier density and the
electric field. Figure \ref{Fig:Eliseev} shows the electroluminescence
peak positions (various symbols) as a function of temperature and
injected current in InGaN single-quantum-well light-emitting-diode,
reported by Eliseev \textit{at al.}.\cite{EliseevPG1997-APL71:569} The
solid curves are the calculated results using
Eqs. (\ref{eq:peak-position}) and (\ref{eq:solve-x}). Excellent
agreement between the theoretical results and the experimental data
was obtained. The values of the parameters used in the calculations
are also listed in Table \ref{Tab:parameters}. From the table, an
interesting observation is that the magnitude of $E_a-E_0$
systematically varies with increasing the injected electrical current
during the measurements. It is known that the increase of the injected
electric current can result in the increase of the carrier
concentration. On the other hand, the increase of the injected
electric current may cause a strength decrease of the effective
electric field in the diode due to the screening out effect of the
electrically injected carriers on the existing huge piezoelectric
field. \cite{TakeuchiT1997-JJAPP236:382, TakeuchiT1998-APL73:1691,
  BernardiniF1997-PRB56:10024, WetzelC2000-PRB61:2159,
  LefebvreP2001-APL78:1252, SalaFD1999-APL74:2002,
  LiQ2002-JJAPP241:L1093} Since the energetic position of $E_a$ is
dependent on both the carrier concentration and the magnitude of the
effective electric field, it is not difficult for one to understand
that the magnitude of $E_a-E_0$ changes with the injected electric
current.

An interesting phenomenon in Fig.~\ref{Fig:Eliseev} is that the
originally scattered electroluminescence peaks in low temperature
region for the different injected currents tend to converge at high
temperatures. Figure \ref{Fig:Eliseev2}(A) depicts the spectral peak
positions of a blue InGaN single-quantum-well light-emitting-diode at
several different temperatures as a function of the injected
current. \cite{EliseevPG2003} At 70 and 100 K, the peak
position increases as the injected current increases. However, when
the temperature $\ge$ 125 K, a plateau region appears. In other
words, the peak position does not change with the injected current in
the region. This indicates that at higher temperatures, the influence
of the magnitude of $E_a-E_0$ on the temperature dependence of the
peak positions will become weaker. This strange behavior can be well
interpreted using Eqs. (\ref{eq:peak-position}) and
(\ref{eq:solve-x}). At high temperatures, the solution of
Eq. (\ref{eq:solve-x}) can be found to be $x\approx(\sigma/k_BT)^2$
and Eq. (\ref{eq:peak-position}) becomes \cite{LiQ-cond-mat-0411128,
  LiQ2001-APL79:1810}
\begin{eqnarray}\label{eq:approx}
  E(T)=E_0-\frac{\alpha T^2}{T+\Theta}-\frac{\sigma^2}{k_BT}.
\end{eqnarray}
Note that $E_a-E_0$ does not appear in the above equation, which
indicates that the peak positions are no longer dependent on $E_a-E_0$
at the high temperatures. Figure \ref{Fig:Eliseev2}(B) shows the
calculated peak positions as a function of $E_a-E_0$ for different
temperatures using Eqs. (\ref{eq:peak-position}) and
(\ref{eq:solve-x}). Compared Fig. 3(A) with (B), it can be concluded
that the behaviors in the experimental data is quantitatively produced
by the theoretical calculations.

In conclusions, the ``S-shaped'' temperature dependence of the
luminescence peak positions, which is a frequently observed phenomenon
in many semiconductors, is successfully modeled using a recently
developed luminescence model for localized state ensemble. The
energetic difference between the quasi-Fermi level ($E_a$) and the
central energy ($E_0$) of the localized state distribution essentially
determines the details of temperature behavior of the luminescence
peaks. The model could be useful for experimentalists to
quantitatively explain their experimental results in temperature
dependent luminescence studies.


This work was supported by the HKU Research Grant
(No. 10203533) and HK RGC grants (No. HKU 7036/03P).


\bibliographystyle{apsrev}%
\bibliography{references,thesis,III-Nitride,mypub,negative}%



\newpage
\vspace{1cm}\noindent\textbf{Figure Captions}

\begin{description}%

\item[FIG. 1.] The ``S-shape'' temperature dependence of luminescence
  peak positions of several semiconductor materials. From the bottom
  to the top: GaInNAs (Ref.~\onlinecite{GrenouilletL2000-APL76:2241}),
  GaInNP (Ref.~\onlinecite{HongYG2003-APL83:5446}), InGaN
  (Ref.~\onlinecite{YuHB2004-CPL21:1323}) and AlGaN
  (Ref.~\onlinecite{BellA2004-JAP95:4670}). The scattered symbols are
  experimental data and the solid lines are the fitting results using
  Eqs.~(\ref{eq:peak-position}) and (\ref{eq:solve-x}).

\item[FIG. 2.] Electroluminescence peak positions (scattered symbols)
  of InGaN light-emitting-diode under different injected currents as a
  function of temperature (from
  Ref.~\onlinecite{EliseevPG1997-APL71:569}). The solid lines are
  calculated results using Eqs.~(\ref{eq:peak-position}) and
  (\ref{eq:solve-x}).

\item[FIG. 3.] (A) Electroluminescence peak positions of InGaN
  light-emitting-diode at several different temperatures as a function
  of injected current (from
  Ref.~\onlinecite{EliseevPG2003}). (B) Calculated spectral
  peak positions for different temperatures as a function of magnitude
  of $E_a\!\!-\!\!E_0$.

\end{description}

\newcommand{\mywidthA}{0.9\linewidth}%
\newcommand{\myheight}{0.9\linewidth}%

\newpage
\begin{figure}[htbp]%
   \centering\includegraphics[height=\myheight]{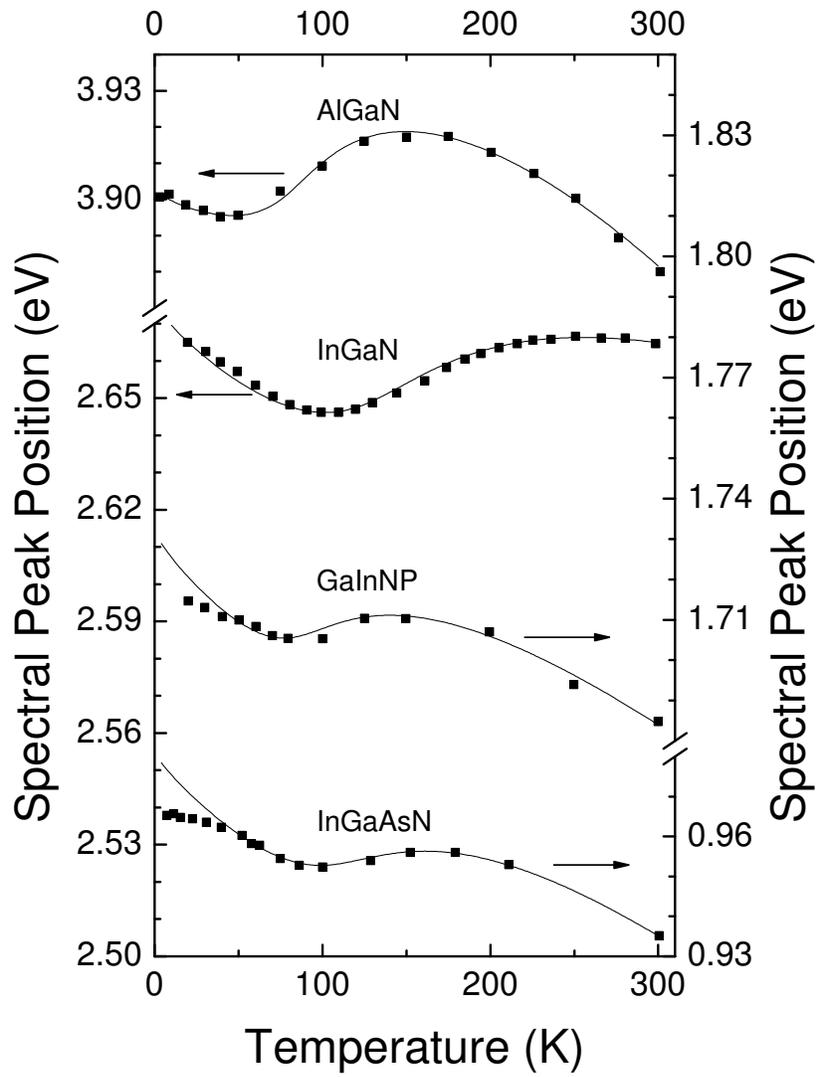}%
   \caption{of 3. Q.~Li, \textit{et al.}}%
   \label{Fig:S-shape}%
\end{figure} 


\newpage
\begin{figure}[htbp]%
   \centering\includegraphics[width=\mywidthA]{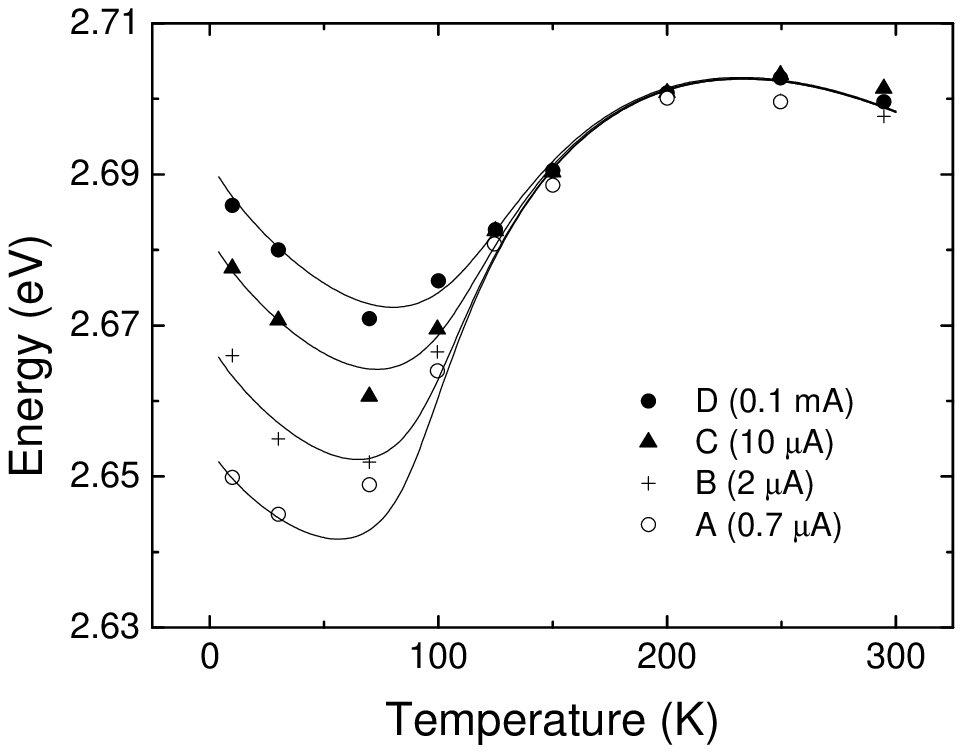}%
   \caption{of 3. Q.~Li, \textit{et al.}}%
   \label{Fig:Eliseev}%
\end{figure}%

\newpage

\begin{figure}[htbp]%
   \centering\includegraphics[clip,bb=148 366 480 666,width=0.2in]{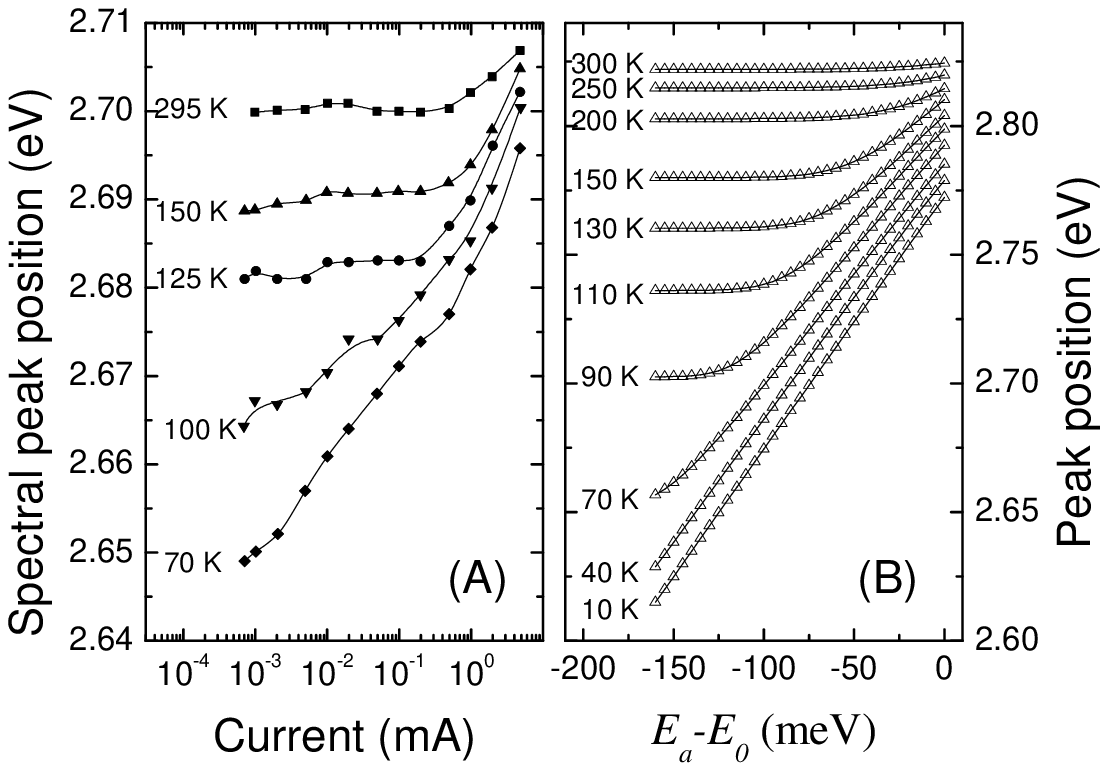}%
   \caption{of 3. Q.~Li, \textit{et al.}}%
   \label{Fig:Eliseev2}%
\end{figure}%

\newpage

\begin{table}[htbp]%
   \caption{Parameters used to fit the temperature dependent energy
   peak positions in Figs.~\ref{Fig:S-shape} and \ref{Fig:Eliseev}.}%
   \label{Tab:parameters}%
   \begin{ruledtabular}\begin{tabular*}{4.1in}{l|cccccc}%
   &$E_0$&$\sigma$&$E_a-E_0$&$\tau_{tr}/\tau_r$&$\alpha$&$\Theta$\\
   &(eV)&(meV)&(eV)&&(meV/K)&(K)\\\hline
   GaInNAs&1.025&25.0&-0.044&0.025&0.44&300\\
   GaInNP&1.763&20.0&-0.031&0.01&0.42&300\\
   InGaN&2.733&30.6&-0.057&0.018&0.35&700\\
   AlGaN&3.989&24.3&-0.086&0.2&0.94&700\\\hline
   A&2.78&31.5&-0.126&0.06&0.5&735\\
   B&2.78&31.5&-0.112&0.04&0.5&735\\
   C&2.78&31.5&-0.098&0.04&0.5&735\\
   D&2.78&31.5&-0.088&0.04&0.5&735
   \end{tabular*}\end{ruledtabular}%
\end{table}%

\end{document}